# Characterization of collective ground states in single-layer NbSe$_2$

Miguel M. Ugeda*[1,2,3], Aaron J. Bradley[1], Yi Zhang[4,5,6], Seita Onishi[1], Yi Chen[1], Wei Ruan[1,7], Claudia Ojeda-Aristizabal[1,8,9], Hyejin Ryu[4], Mark T. Edmonds[1,10], Hsin-Zon Tsai[1], Alexander Riss[1,11], Sung-Kwan Mo[4], Dunghai Lee[1], Alex Zettl[1,8,12], Zahid Hussain[4], Zhi-Xun Shen[5,13] and Michael F. Crommie*[1,8,12]

[1]Department of Physics, University of California at Berkeley, Berkeley, California 94720, USA.

[2]CIC nanoGUNE, 20018 Donostia-San Sebastian, Spain.

[3]Ikerbasque, Basque Foundation for Science, 48011 Bilbao, Spain.

[4]Advanced Light Source, Lawrence Berkeley National Laboratory, Berkeley, California 94720, USA.

[5]Stanford Institute for Materials and Energy Sciences, SLAC National Accelerator Laboratory, Menlo Park, California 94025, USA.

[6]National Laboratory of Solid State Microstructures, School of Physics, Collaborative Innovation Center of Advanced Microstructures, Nanjing University, Nanjing 210093, China.

[7]State Key Laboratory of Low Dimensional Quantum Physics, Department of Physics, Tsinghua University, Beijing 100084, China.

[8]Materials Sciences Division, Lawrence Berkeley National Laboratory, Berkeley, California 94720, USA.

[9]Department of Physics & Astronomy, California State University Long Beach, Long Beach, California 90840, USA.

[10]School of Physics and Astronomy, Monash University, Clayton,Victoria 3800, Australia.

[11]Institute of Applied Physics, Vienna University of Technology, 1040Wien, Austria.

[12]Kavli Energy NanoSciences Institute at the University of California Berkeley and the Lawrence Berkeley National Laboratory, Berkeley, California 94720, USA.

[13]Geballe Laboratory for Advanced Materials, Departments of Physics and Applied Physics, Stanford University, Stanford, California 94305, USA.

* *Corresponding authors:* mmugeda@berkeley.edu *and* crommie@berkeley.edu

# Abstract

Layered transition metal dichalcogenides (TMDs) are ideal systems for exploring the effects of dimensionality on correlated electronic phases such as charge density wave (CDW) order and superconductivity. In bulk NbSe$_2$ a CDW sets in at $T_{CDW}$ = 33 K and superconductivity sets in at $T_c$ = 7.2 K. Below $T_c$ these electronic states coexist but their microscopic formation mechanisms remain controversial. Here we present an electronic characterization study of a single 2D layer of NbSe$_2$ by means of low temperature scanning tunneling microscopy/spectroscopy (STM/STS), angle-resolved photoemission spectroscopy (ARPES), and electrical transport measurements. We demonstrate that 3x3 CDW order in NbSe$_2$ remains intact in 2D. Superconductivity also still remains in the 2D limit, but its onset temperature is depressed to 1.9 K. Our STS measurements at 5 K reveal a CDW gap of $\Delta$ = 4 meV at the Fermi energy, which is accessible via STS due to the removal of bands crossing the Fermi level for a single layer. Our observations are consistent with the simplified (compared to bulk) electronic structure of single-layer NbSe$_2$, thus providing new insight into CDW formation and superconductivity in this model strongly-correlated system.

Many-body electronic ground states can be quite sensitive to the spatial dimensions of a material[1, 2, 3, 4]. In transition metal dichalcogenide materials, for example, significant differences are expected in charge density wave and superconducting (SC) phases as dimensionality is reduced from a bulk 3D material to a single-layer 2D material[5, 6, 7]. $NbSe_2$ is a model system in this regard, as it has been predicted to preserve its CDW order in the single-layer limit, although with significantly shortened CDW wavevector ($\mathbf{q}_{CDW}$)[5]. A metal to semi-metal transition has also been predicted for $NbSe_2$ when it is reduced to the 2D limit [5, 6, 8]. Previous work has shown that the superconducting transition temperature for $NbSe_2$ decreases from its bulk value of $T_C = 7.2$ K as the thickness is reduced[9, 10, 11], but no experimental studies have yet been performed that explore the interplay between $NbSe_2$ superconductivity and CDW formation in the extreme 2D limit.

CDW behavior in the 3D limit of $NbSe_2$, on the other hand, has been well studied, although the origin of the bulk CDW transition at $T_{CDW} = 33K$ remains controversial. Recent experiments suggest that electron-phonon coupling plays a dominant role in triggering the CDW phase[12, 13, 14, 15, 16, 17, 18], thus casting doubt on mechanisms involving Fermi-surface nesting[19, 20, 21, 22] and saddle-point singularities[23, 24]. Critical experimental parameters associated with the CDW, namely the magnitude and location of the energy gap also remain unclear. Previous STS measurements on bulk $NbSe_2$ in the CDW phase have revealed an unexpectedly large gap-like structure having width $2\Delta \sim 70$ meV[25, 26]. In contrast, low-temperature ARPES measurements performed on bulk $NbSe_2$ show an anisotropic CDW energy gap at $E_F$ with a width of only a few meV[15]. In general, it is expected that a weak-coupling CDW derived from Fermi surface nesting should open a small gap on the Fermi surface, while a strong coupling CDW caused by electron-phonon coupling should open a larger gap away from the Fermi surface.

Here we report new measurements on the electronic structure of single-layer $NbSe_2$ using a combination of STM/STS, ARPES, and electronic transport techniques. This allows us to directly probe the electronic ground state of $NbSe_2$ in the 2D limit, enabling interrogation of the effects of dimensionality and interlayer coupling in this layered material. We observe a reduction in the number of Fermi-level-crossing bands from three (for bulk) to one in the single-layer limit. Despite this change in electronic structure, CDW order in single-layer $NbSe_2$ remains unchanged compared to the bulk case. The simplified bandstructure of single-layer $NbSe_2$, however, allows unprecedented access to the $NbSe_2$ CDW energy gap via STS (observed here to be $2\Delta = 8$ meV centered at $E_F$). Superconductivity in $NbSe_2$, on the other hand, is significantly suppressed in the 2D limit, showing an onset of superconducting fluctuations at 1.9 K and a broadened superconducting transition with a midpoint at 0.65 K consistent with a Kosterlitz-Thouless transition. The one Nb antibonding band[5] that remains in the Fermi surface of single-layer $NbSe_2$ thus appears to play a critical role in the $NbSe_2$ CDW state, while the absent bands have a stronger influence on $NbSe_2$ superconductivity[15]. These results provide a new window into the electronic structure of single-layer $NbSe_2$ and help to clarify the longstanding debate over $NbSe_2$ CDW formation.

Our experiments were carried out on high-quality submonolayer $NbSe_2$ films grown on epitaxial bilayer graphene (BLG) on 6H-SiC(0001), as sketched in Fig. 1a. The large-scale STM image in Fig. 1b shows the typical morphology of our single-layer $NbSe_2$ samples. Black regions correspond to the BLG substrate while the $NbSe_2$ layer is purple. The temperature dependence of the electronic ground state of single-layer $NbSe_2$ was measured via STM and electrical transport. Fig. 1c-e show STM topographic data for selected temperatures from T = 45 K to T = 5 K. At T = 45 K, well above the critical transition temperature for bulk $NbSe_2$ ($T_{CDW} = 33$ K), only the undistorted crystal structure is observed (Fig. 1c). At a lower temperature of T = 25 K, weak and spotty signatures of a superlattice are apparent (Fig. 1d). Here small CDW patches surrounded by non CDW regions can be seen. This is reminiscent of STM images of bulk NbSe$_2$ at temperatures close to the CDW transition temperature.[17] At T = 5 K, the 3 x 3 CDW superlattice is fully and uniformly developed for single-layer NbSe$_2$ (Fig. 1e). Fig. 2 shows the temperature-dependent electrical resistance of single-layer NbSe$_2$ on BLG, acquired using a 4-probe low-excitation dc method (see SI). A downturn in the resistance begins at T = 1.9 K, indicative of the onset of superconducting fluctuations, while the superconducting transition midpoint is at 0.65 K and the zero resistance point at 0.46 K, as shown inset. This data indicate that the trend of reduced superconducting transition (T$_c$) with decreasing layer number in NbSe$_2$[11, 12] continues down to the single layer limit.

Fig. 1e shows that the 3 x 3 CDW superlattice is aligned with the 1 x 1 atomic arrangement for single-layer NbSe$_2$, similar to what has been seen previously in STM images of bulk NbSe$_2$[16, 17, 27]. The 3 x 3 superlattice remains unchanged in our STM images regardless of the orientation between the NbSe$_2$ layer and the BLG (see STM and LEED data in the SI). This rules out the possibility that the 3 x 3 superlattice observed here in single-layer NbSe$_2$ is a moiré pattern (moiré patterns have been observed in MoSe$_2$/BLG[28]). A 3 x 3 moiré pattern formation requires a quasi-commensurate match between overlayer and substrate atomic lattices, which does not occur here since NbSe$_2$ has a much larger unit cell than graphene. The rotational disorder we find in single-layer NbSe$_2$ with respect to the BLG substrate indicates weak coupling between them (similar weak coupling has been demonstrated for epitaxial graphene grown on different metal substrates[29]).

We experimentally determined the electronic structure of single-layer NbSe$_2$ through a combination of STS and ARPES. Fig. 3a shows a typical STM dI/dV spectrum of single-layer NbSe$_2$ taken over a large bias range. In the positive bias (empty state) region the most pronounced feature is the peak labeled C$_1$ at V$_s$ = 0 .5 V. For negative bias (filled states) we observe a very shallow asymmetric peak near $V_s = -0.2$ V (labeled $V_1$), below which the local density of states (LDOS) flattens out and does not rise again until the peak labeled $V_2$ at $V_s = -0.8$ V. This behavior is very different from previous STS results obtained for bulk $NbSe_2$, which show much higher LDOS in the region -0.8 V < V < 0 V[17]. Fig. 3b shows the ARPES data obtained from the same type of single-layer $NbSe_2$ sample as in Fig. 3a using *sp*-mixed polarized light. The measured electronic dispersion is energetically aligned with the dI/dV spectrum for comparison. Due to rotational misalignment within the $NbSe_2$ layer, the ARPES dispersion reflects a mixture of states with dominant intensity coming from the high symmetry directions Γ-M and Γ-K. Two dispersive features can be seen crossing $E_F$ (highlighted by the orange dashed lines). The ARPES spectrum also shows an energy gap over the range -0.8 eV < E < -0.3 eV, and below this several dispersive bands merge at the Γ-point. In comparison, the ARPES spectrum for 5ML–$NbSe_2$ (see SI) displays an additional Nb-derived band crossing the same energy range that is gapped for single-layer $NbSe_2$.

This experimental data is consistent with the changes in band structure that are predicted to occur when $NbSe_2$ is thinned down to a single layer.[5] The bulk $NbSe_2$ band structure (Fig. 3c) has three bands crossing $E_F$ (two Nb-derived bands and one S-derived band), and exhibits no bandgap throughout our experimental energy range. Single-layer $NbSe_2$, on the other hand, has a much simpler predicted electronic structure (Fig. 3d) that consist of just one Nb-derived band crossing $E_F$ away from the Γ-point, as well as a band gap from -0.4 eV to -0.8 eV and valence bands below that. These predicted features can be seen in our experimental ARPES dispersion (Fig. 3b) which shows that (i) states right below $E_F$ are located away from the Γ-point, (ii) there exists an energy gap between -0.3 eV and -0.8 eV, and (iii) there exists several valence bands below -0.8 eV. Our dI/dV spectroscopy (Fig. 3a) is also consistent with the predicted single-layer band structure. For example, the peak at $C_1$ is consistent with the high DOS associated with the top of the Nb-derived band at Γ above $E_F$. We can also associate the shallow peak at $V_1$ with the Nb band right below $E_F$. The flat dI/dV region seen between $V_1$ and $V_2$ is consistent with the predicted bandgap (the fact that the dI/dV does not reach zero is due to residual tunneling into the BLG substrate), and the peaks at $V_2$ and $V_3$ are consistent with the lower-energy valence bandstructure[5, 30]. Overall, these features are very different from previous STS spectra obtained for bulk $NbSe_2$, which show much higher LDOS below $E_F$ and no signs of a bandgap in the filled-state regions[17]. This is due to the additional electronic bands and resulting DOS expected in this energy range for bulk $NbSe_2$ as compared to single-layer $NbSe_2$ (Figs. 3c-d). This is also reflected in the additional band that can be seen in the ARPES spectrum of 5ML–$NbSe_2$ (SI), which corresponds to a Nb bonding band (the blue band in Figs. 3c-d). The Se-derived band (green band in Figs. 3c-d) is typically not observed in ARPES due to the high $k_z$ dispersion.

In order to better understand the collective ground states of single-layer $NbSe_2$, we experimentally probed its low-energy electronic structure near $E_F$ via STS. Fig. 4 shows a typical low-bias dI/dV spectrum obtained for single-layer $NbSe_2$. This spectrum exhibits a striking energy gap feature centered at $E_F$ that is not present in the calculated bandstructure of single-layer $NbSe_2$ (Fig. 3d). This feature, which is also not observed experimentally in bulk $NbSe_2$[16, 25, 26], exhibits a sharp dip at $E_F$ bounded by two narrow peaks that sit on top of an asymmetric background. The dip at $E_F$ does not reach all the way to zero, suggesting that it is not a full gap in the electronic structure. These STS features (including both wide-bias and low-bias spectra) were observed consistently for hundreds of dI/dV curves measured on several samples using a variety of different tip apexes. However, the width ($2\Delta$) of the low-bias gap feature was found to exhibit some spatial variation, likely due to heterogeneity induced by the presence of defects at the $NbSe_2$/graphene interface (see Fig. 5c). Statistical analysis of data (257 curves) obtained at many different locations for T = 5 K yields an average gap value of $\Delta = 4.0$ mV with a standard deviation of 1 mV (the gap magnitude is defined as half the energy distance between the two peaks bracketing the gap). This value is in agreement with anisotropic gap-opening observed previously by low-temperature ARPES[15] at the Fermi surface near the K point in the band that is predicted to remain in the 2D single-layer limit of NbSe$_2$ (orange band in Fig. 3d).

In order to better understand the origin of the E$_F$ gap feature of Fig. 4, we performed spatially-resolved dI/dV mapping of single-layer NbSe$_2$ at different bias voltages near E$_F$. Figs. 5a and 5b show dI/dV conductance maps of the same region taken at bias voltages outside (V$_b$ = -26 mV) and inside (V$_b$ = 3 mV) the low-bias gap region (Fig. 5c shows STM topography for this same region). The dI/dV conductance map taken at an energy outside of the gap (Fig. 5a) clearly shows the 3 x 3 CDW pattern, but the conductance map measured at an energy inside of the gap (Fig. 5b) shows no sign of the CDW. We Fourier analyzed dI/dV maps taken at different voltages in order to obtain a more quantitative understanding of the energy dependence of CDW electronic features. Fig. 5d shows the relative intensity of the resulting 3 x 3 peaks (I$_{3x3}$) in the FFTs normalized to the intensity of the 1 x 1 Bragg peaks (I$_{Bragg}$). This ratio (I$_{3x3}$/I$_{Bragg}$(E)) is a measure of the strength of the CDW modulation on the density of states. As seen in Fig. 5d, I$_{3x3}$/I$_{Bragg}$ has large amplitude at energies far from E$_F$, but shows a decrease of nearly two orders of magnitude very close to E$_F$. Fig. 5e shows a higher-resolution plot of the dip in I$_{3x3}$/I$_{Bragg}$ at E$_F$ (boxed region in Fig. 5d). The decrease in the ratio I$_{3x3}$/I$_{Bragg}$ is seen to follow the energy dependence of the low-bias gap-feature observed in dI/dV spectroscopy. This correlation of CDW intensity with gap energy dependence suggests that the gap is the result of CDW order. In particular, the diminishing CDW intensity for in-gap energies suggests that the residual LDOS within the gap arises from a portion of the Fermi surface that is not gapped by CDW order.

Our results allow us to draw some conclusions regarding different models of CDW formation in NbSe$_2$. First, we rule out recently proposed dimensionality effects on the CDW phase that were predicted for single-layer NbSe$_2$ but that are not observed here. This includes a predicted reduction of the CDW wave vector in the 2D limit[5]. We also rule out proposed Fermi surface nesting mechanisms[17, 19, 20, 22] involving the inner pockets around Γ and K (blue band in Fig.3c) since these bands are not present in the single-layer limit and the CDW remains unchanged. Since saddle-point-based mechanisms of CDW formation[21, 24] involve the Fermi pockets that are still present in the single-layer limit, we cannot rule them out based on fermiology. However, these mechanisms predict a CDW gap centered at an energy tens of meV away from $E_F$, and thus are inconsistent with our observation that the CDW gap is centered at $E_F$.

Our data also brings out some puzzling features concerning the CDW in NbSe$_2$. On the one hand, our observations that the gap is tethered to the Fermi energy, that is rather small, and that it correlates with the CDW amplitude are all consistent with the notion that the gap is the result of Fermi surface nesting. On the other hand, our observation that the CDW modulation is observed in the LDOS at biases far exceeding the gap edges suggests that the CDW order is not a weak coupling phenomenon arising from Fermi surface nesting. This dual nature of the CDW gap imposes stringent constraints on any future theory of CDW formation in NbSe$_2$.

The suppression of the superconducting onset temperature in single-layer NbSe$_2$ is consistent with the trend previously observed[9, 10, 11], namely that the superconducting transition temperature decreases with sample thinning. Possible reasons for this include enhancement of thermally driven superconducting phase fluctuations, as well as weakening of the strength of Cooper pairing in 2D. Additional factors that may contribute to this weakening are the reduced screening of the Coulomb interaction[10] and the reduction of DOS at $E_F$ for single-layer NbSe$_2$ arising from band reduction. It is likely that the electronic bands that are absent in single-layer NbSe$_2$ play a key role for thicker NbSe$_2$ films[15].

**Methods**

Single-layer NbSe$_2$ was grown by molecular beam epitaxy (MBE) on epitaxial BLG on 6H-SiC(0001) at the HERS endstation of beamline 10.0.1, Advanced Light Source, Lawrence Berkeley National Laboratory (the MBE chamber had a base pressure of ~ $2 \times 10^{-10}$ Torr). We used SiC wafers with two different resistivities, ρ ~ 300 Ωcm (STM and ARPES experiments) and ρ > $10^8$ Ωcm (transport and STM experiments). The morphology of the single-layer NbSe$_2$ was the same in both cases. The epitaxial BLG substrate was prepared by following the procedure detailed in refs. [31] and [28]. High purity Nb and Se were evaporated from an electron-beam evaporator and a standard Knudsen cell, respectively. The flux ratio of Nb to Se was controlled to be ~ 1:30. The growth process was monitored by in-situ RHEED and the growth rate was ~17 minutes per monolayer. During the growth the substrate temperature was kept at 600 K, and after growth the sample was annealed to 670 K. Low energy electron diffraction (LEED) patterns were routinely taken after the growth to determine the quality of the sample (See SI). Subsequent characterization by ARPES and core-level spectroscopy was performed in the analysis chamber (base pressure ~$3 \times 10^{-11}$ Torr) of the 10.0.1 beamline. To protect the film from contamination and oxidation during transport through air to the UHV-STM chamber, a Se capping layer with a thickness of ~10 nm was deposited on the sample surface after growth. For subsequent STM and transport experiments the Se capping layer was removed by annealing the sample to ~ 600 K in the UHV STM system for 30 minutes. STM imaging and STS experiments were performed in a home-built UHV-STM operated at T = 5 K. STM differential conductance (dI/dV) spectra were measured at 5 K using standard lock-in techniques. To avoid tip artifacts the STM tip was calibrated by measuring reference spectra on the graphene substrate[28]. STM/STS data were analyzed and rendered using WSxM software[32].

Four-probe contacts for transport measurements were placed on the sample by electron beam evaporation of 6.5 nm Ti and 100 nm Au through a shadow mask onto exposed graphene portions of the $NbSe_2$ sample (with a Se capping layer, see SI). The Se capping layer was then removed in UHV and resistance measurements were subsequently performed either *in-situ* inside the STM chamber, or in a Quantum Design PPMS dilution refrigerator (employing a rapid transfer from the UHV chamber to the inert gas/vacuum PPMS chamber). Measurements were made using a Keithley 2602A source-measure unit and a Keithley 181 nanovoltmeter with a current bias of 100nA. At each temperature point the voltage was measured at both 100nA and -100nA to remove thermoelectric voltages induced by the temperature difference between the sample and measurement electronics. The heating rate during measurement ranged from 1 K/hr to 60 K/hr.

**Acknowledgments:**

Research supported in part by the Director, Office of Energy Research, Materials Sciences and Engineering Division, of the U.S. Department of Energy (DOE), under grant DE-AC02-05CH11231 supporting the sp2-bonded Materials Program (STM imaging and transport), and by the National Science Foundation under award # DMR-1206512 (STS spectroscopic analysis). Work at the ALS is supported by DOE BES under Contract No. DE-AC02-05CH11231. H.R. acknowledges support from Max Planck Korea/POSTECH Research Initiative of NRF, Korea. M.T.E. is supported by the ARC Laureate Fellowship project (FL120100038). A.R. acknowledges fellowship support by the Austrian Science Fund (FWF): J3026-N16.

**Author contributions:**

M.M.U. and A.J.B. conceived the work and designed the research strategy. M.M.U., A.J.B., Y.C., W.R., M.T.E. measured and analyzed the STM/STS data. Y.Z., H.R. and S-K.M. performed the MBE growth and ARPES and LEED characterization of the samples. S.O., C.O.A., M.M.U. and Y.C. carried out the transport and experiments. H.T. and A.R. helped in the experiments. D.L. participated in the interpretation of the experimental data. Z.H. and Z-X.S. supervised the MBE and sample characterization. A.Z. supervised the transport measurements. M.F.C. supervised the STM/STS experiments. M.M.U. wrote the paper with help from M.F.C. and A.Z.. M.M.U. and M.F.C. coordinated the collaboration. All authors contributed to the scientific discussion and manuscript revisions.

**Additional information**

Supplementary information is available in the online version of the paper. Reprints and permissions information is available online at www.nature.com/reprints.

**Competing financial interests**

The authors declare no competing financial interests.

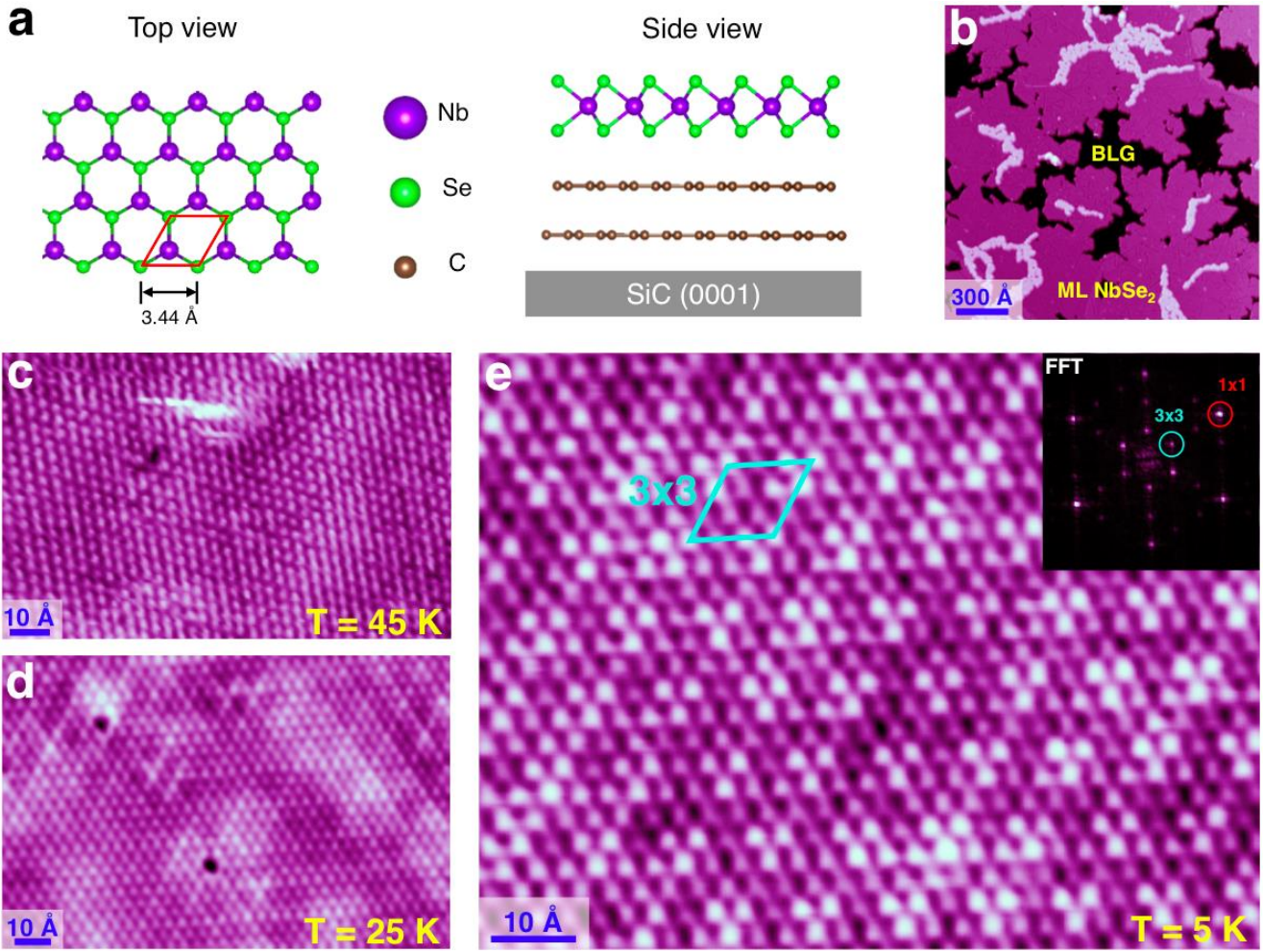

**Figure 1. Structure of single-layer NbSe₂ on bilayer graphene. a,** Top and side view sketches of single-layer NbSe$_2$, including the substrate. **b,** Large-scale STM image of 0.9 ML of NbSe$_2$/BLG ($V_s$ = - 100 mV, $I_t$ = 3 nA, T = 5 K). Atomically-resolved STM images of single-layer NbSe$_2$ are shown for different temperatures: **c,** T = 45 K ($V_s$ = + 95 mV, $I_t$ = 53 pA), **d,** T = 25 K ($V_s$ = - 120 mV, $I_t$ = 70 pA) and **e,** T = 5 K ($V_s$ = - 4 mV, $I_t$ = 50 pA). The FFT of the STM image of **e** is shown in the inset.

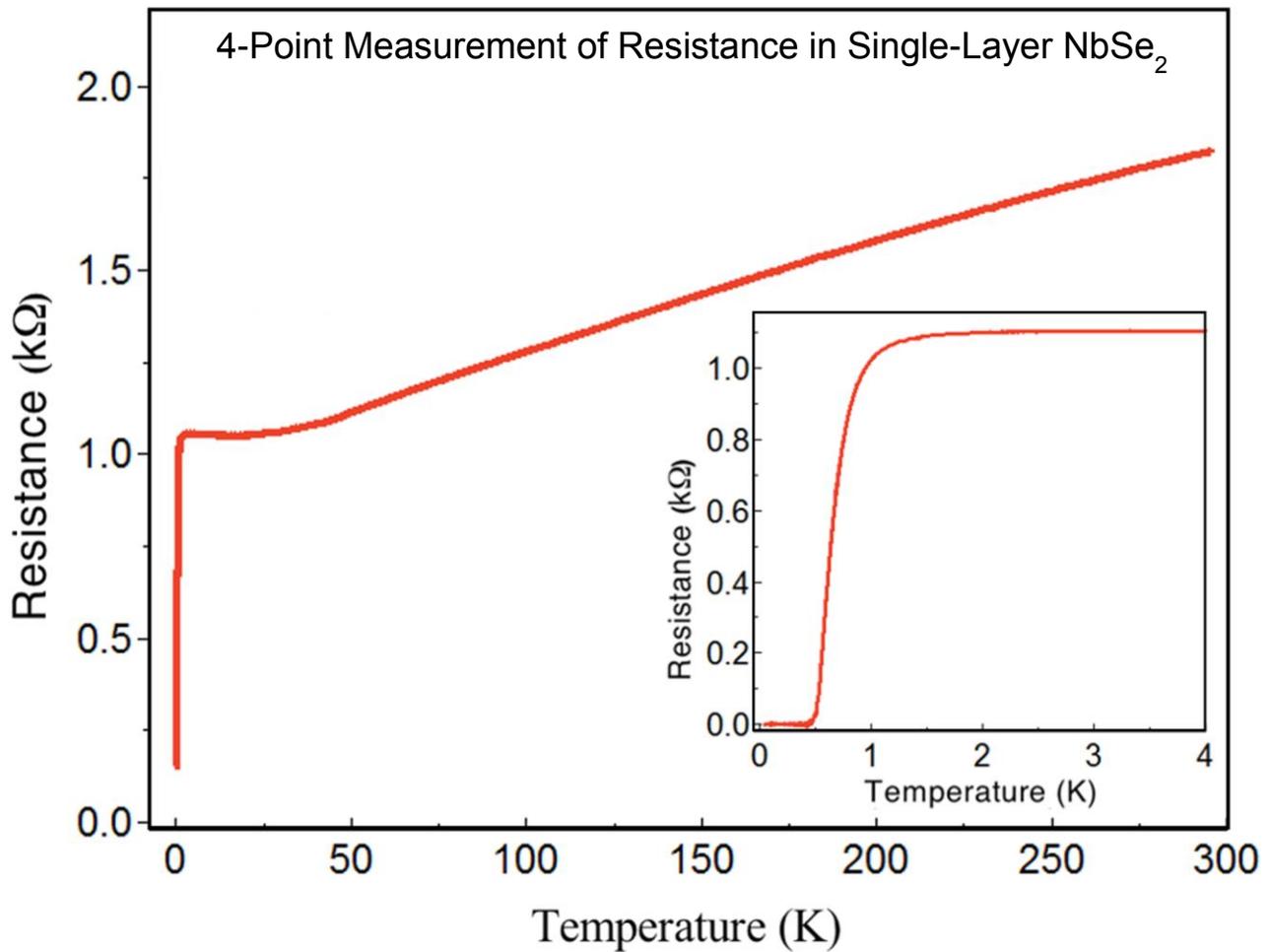

**Figure 2. Superconductivity in single-layer NbSe₂ on bilayer graphene.** 4-point probe measurement of the temperature-dependent resistance of single-layer $NbSe_2$/BLG on insulating SiC(0001). Inset shows expanded view of R vs. T between 0.05 K and 4 K.

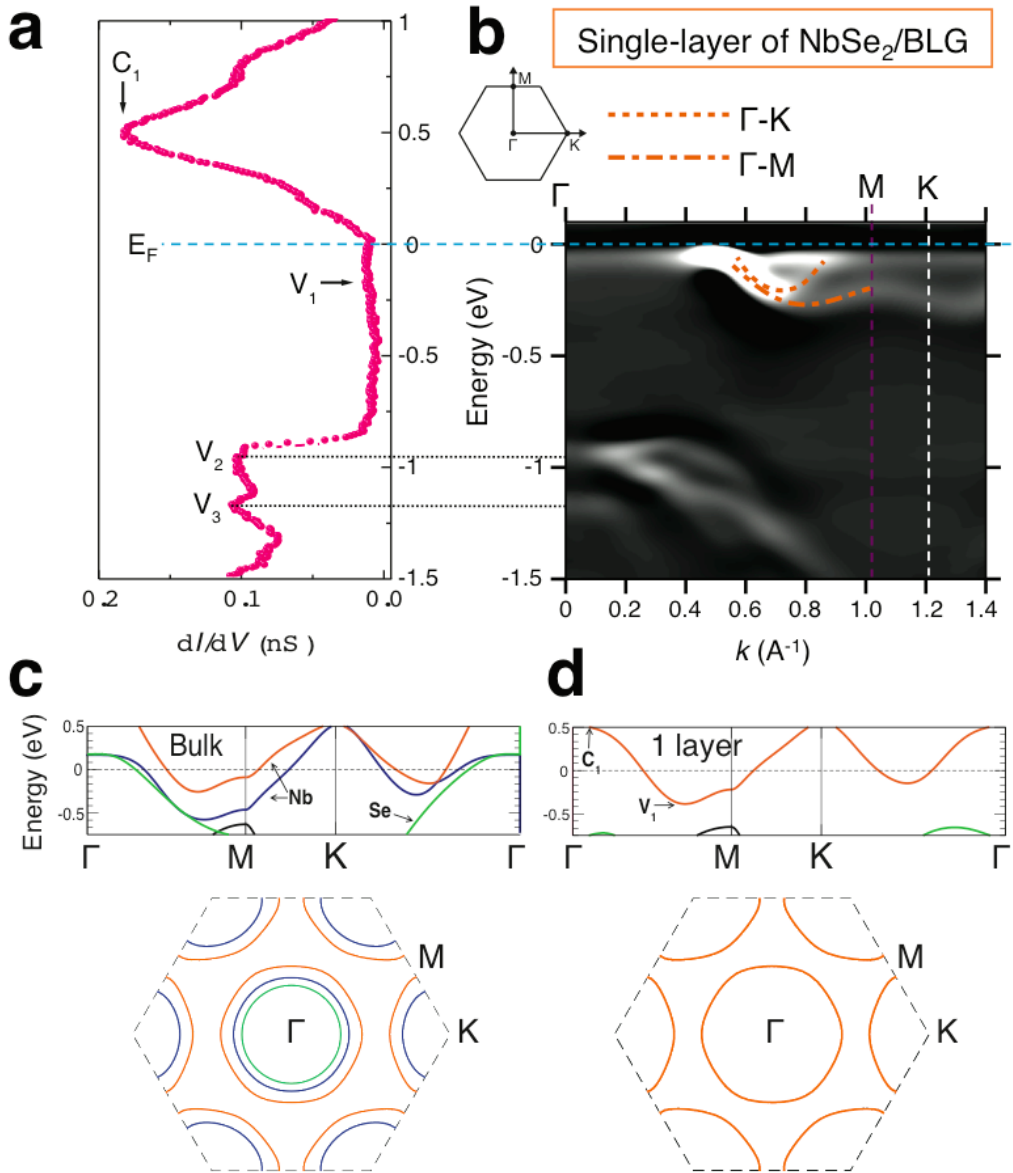

**Figure 3. Electronic structure of single-layer NbSe$_2$ on bilayer graphene. a,** Wide-bias STM dI/dV spectrum acquired on single-layer NbSe$_2$/BLG showing several electronic features: V$_{1-3}$ in the filled states and C$_1$ in the empty states (f = 403 Hz, I$_t$ = 100 pA, V$_{rms}$ = 5 mV, T = 5 K). **b,** Second-derivative *sp*-polarized ARPES dispersion of single-layer NbSe$_2$/BLG (T = 300 K) aligned in energy with the STM dI/dV curve shown in **a**. The dispersion exhibits angular integration due to the intrinsic rotational misalignment of NbSe$_2$ domains on the BLG substrate. The dotted and dot-dashed orange curves indicate bands from the Γ-K and Γ-M directions, respectively. The different predicted bandstructures and Fermi surfaces of bulk NbSe$_2$ and single-layer NbSe$_2$ are shown in **c** and **d**, respectively. The calculated bandstructures (DFT) have been adapted from ref [5].

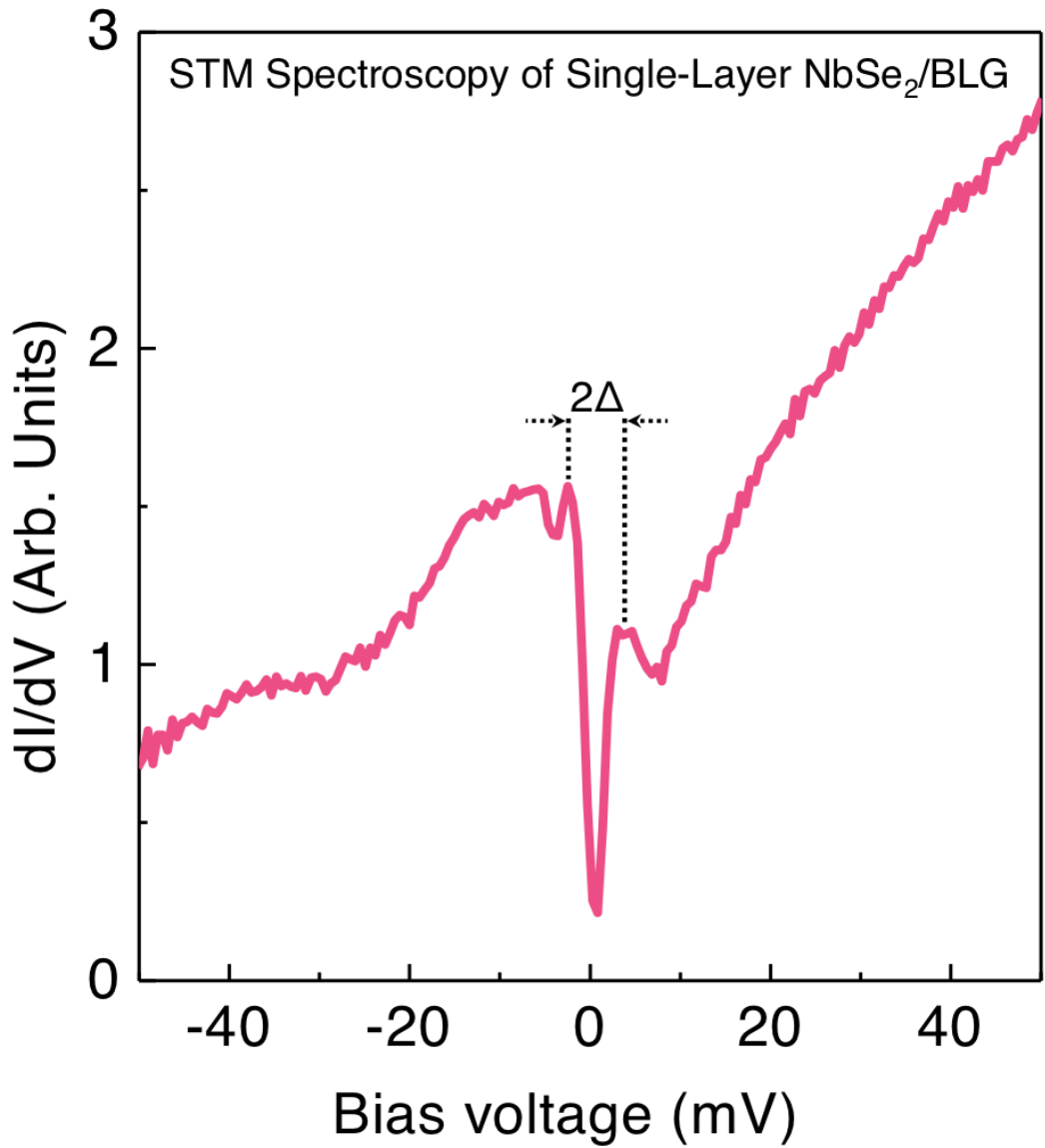

**Figure 4. CDW gap of single-layer NbSe$_2$.** Low-bias STM dI/dV spectrum acquired on single-layer NbSe$_2$/BLG showing the CDW gap ($\Delta$) at E$_F$ (f = 871 Hz, I$_t$ = 100 pA, V$_{rms}$ = 0.6 mV, T = 5 K). The dashed lines indicate the position of the CDW coherence peaks.

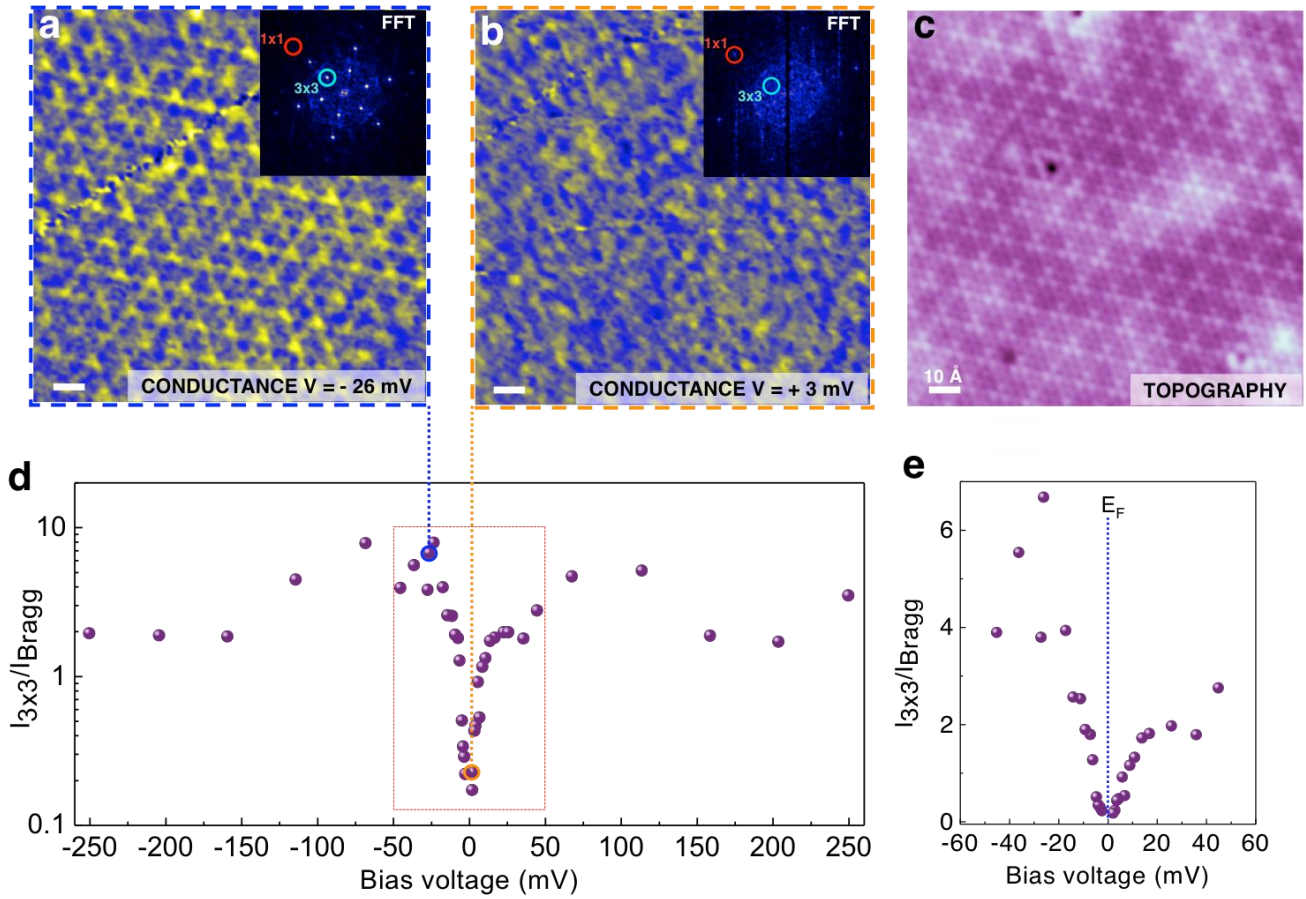

**Figure 5. Spatially and energetically resolved CDW phase in single-layer NbSe$_2$.** Experimental conductance maps taken at (**a**) $V_s$ = - 26 mV and (**b**) $V_s$ = + 3 mV (f = 871 Hz, $I_t$ = 40 pA, $V_{rms}$ = 0.6 mV, T = 5 K). The FFT of each conductance map is shown as an inset, allowing identification of Bragg (1x1) and CDW (3x3) peaks. **c**, STM topograph of the same region where the conductance maps in **a** and **b** were acquired ($V_s$ = - 17 mV, $I_t$ = 40 pA, T = 5 K). **d**, Logarithmic plot of the intensity of the 3 x 3 peaks ($I_{3x3}$) in the FFT of the different conductance maps normalized by the Bragg peak intensity ($I_{Bragg}$) as a function of bias voltage. **e**, Zoom-in of the boxed region in **d**.

# Supplementary Materials for

# Characterization of collective ground states in single-layer 2H-NbSe$_2$

Miguel M. Ugeda*, Aaron J. Bradley, Yi Zhang, Seita Onishi, Yi Chen, Wei Ruan, Claudia Ojeda-Aristizabal, Hyejin Ryu, Mark T. Edmonds, Hsin-Zon Tsai, Alexander Riss, Sung-kwan Mo, Zahid Hussain, Dunghai Lee, Alex Zettl, Zhi-Xun Shen and Michael F. Crommie*.

*Correspondence to: mmugeda@berkeley.edu and crommie@berkeley.edu.

# 1. ARPES characterization of the monolayer and 5ML- NbSe$_2$

The *in-situ* angle-resolved photoemission spectroscopy (ARPES) measurements were performed at the HERS endstation of beamline 10.0.1 at the Advanced Light Source, Lawrence Berkeley National Laboratory. Samples were cooled to ~60K during measurement. The size of the beam spot on the sample was ~150 μm × 200 μm, and the photon energy was 50 eV with energy and angular resolution of 25 meV and 0.1º, respectively. Figure S1a shows the geometry of the ARPES measurement. Two types of photon polarizations were used. In one polarization the photon electric field was perpendicular to the plane of incidence (i.e. *s*-polarization). In the other polarization, the photon electric field was 20° out of the plane of incidence (a mixture of *s*-polarization and *p*-polarization referred to as *sp-mixed* polarization).

Figure S1b shows the hexagonal 2D Brillouin zone (black dotted hexagon) and a schematical drawing of the Fermi surface (orange dotted lines) for single-layer NbSe$_2$, with a near-hexagon pocket at the Γ point and a near- triangle pocket at each of the six K points. Figures S1c and S1d show Fermi surface maps taken with *s*-polarized photons and *sp*-mixed polarized photons, respectively. We found that the NbSe$_2$ films contain randomly rotated domains with two dominant, preferred orientations. The two dominant orientations, separated by a 30º difference, are depicted in figures S1c and S1d as dotted orange hexagons, in good overall agreement with the Fermi surface mapping data. The randomly rotated domains were also observed in the scanning tunneling microscopy (STM) measurements (see fig. S2).

Even though the ARPES spectra along Γ-M and Γ-K directions were mixed due to the randomly rotated domains, we still can distinguish the Γ-M and Γ-K bands by using different photon polarizations[1]. Figs. S1e and S1g are the ARPES spectra of a monolayer NbSe$_2$ film taken with *s*- polarization and *sp-mixed* polarization, respectively. The corresponding second-derivative spectra are provided for enhanced visibility (figures S1f and S1h). Using *s*-polarized photons, only the Γ-M band (depicted by the orange dotted line) can be observed in figs. S1e and S1f; using *sp-mixed* polarized photons, both the Γ-M band and the Γ-K band (depicted by dotted and dot-dashed orange lines, respectively) can be observed in figs. S1g and S1h. Importantly, we found only one band crossing the Fermi level for single-layer NbSe$_2$, which agrees well with recent theoretical calculations (fig. S1k).[2]

We also grew and characterized thicker NbSe$_2$ films. Unfortunately, we noticed that the quality of few-layer NbSe$_2$ reduces gradually with the number of layers. Figs. S1i and S1j are the ARPES spectra and second-derivative spectra of a 5 ML NbSe$_2$ film. The ARPES spectra of this 5 ML NbSe$_2$ film became significantly more blurred than those of the single layer. However, we still can clearly observe a second band crossing the Fermi level in the second-derivative spectrum (depicted by the blue curve in figs. S1i and S1j). Comparison of this data with theoretical calculations for the bulk NbSe$_2$ band structure[2] allows us to conclude that the additional band in 5 ML NbSe$_2$ is the Nb-derived band of bulk NbSe$_2$, shown in blue in fig. S1k. We did not observe the theoretically predicted Se-derived band from bulk NbSe$_2$ shown in green in fig. S1k. However, our ARPES results indicate that single-layer NbSe$_2$ has distinctly different band structure from 5 ML NbSe$_2$.

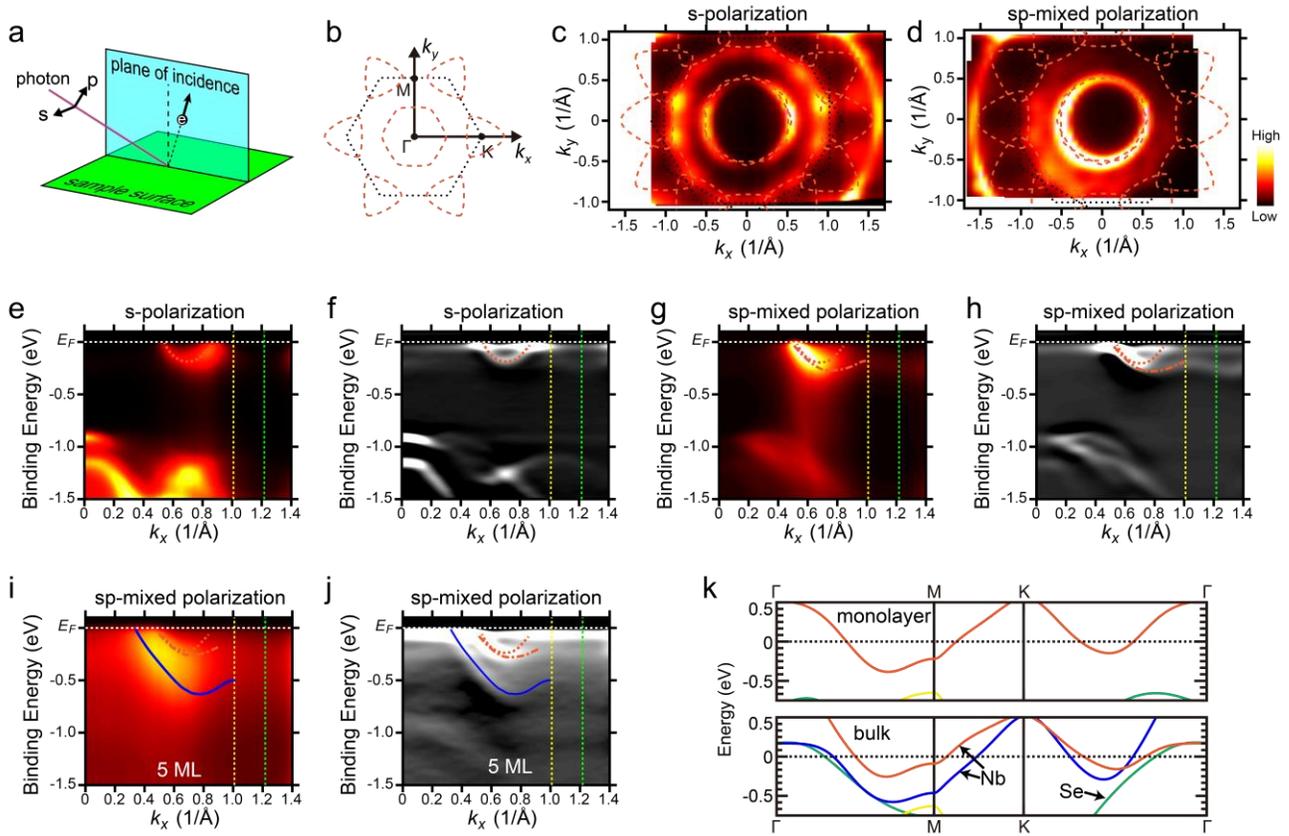

**Figure S1: ARPES spectra of NbSe$_2$ films on BLG. a,** Schematic drawing of the geometric setting in ARPES measurement. **b,** 2D Brillouin zone (dotted hexagon) and Fermi surface sketch map (orange dotted lines) of single-layer NbSe$_2$. **c,** and **d,** Fermi surface mapping of single-layer NbSe$_2$ film by using *s*-polarization photon and *sp*-mixed polarization photon, respectively; the dotted lines depict the Brillouin zone (black) and Fermi surface (orange) for the two dominantly rotated lattice orientations, respectively. **e,** ARPES spectra and **f,** the second-derivative spectra of monolayer NbSe$_2$ film taken with *s*-polarization photon. **g,** ARPES spectra and **h**, the second-derivative spectra of single-layer NbSe$_2$ film taken with *sp*-mixed polarization photon. The dotted and dot-dashed orange curves indicate bands from the Γ-K and Γ-M directions, respectively. **i,** ARPES spectra and **j,** the second-derivative spectra of 5 ML NbSe$_2$ film taken with *sp*-mixed polarization photon. **k)** Theoretical calculations of the band structures for monolayer and bulk NbSe$_2$ (ref. 2).

## 2. Rotational misalignment of single-layer NbSe$_2$ on bilayer graphene

Consistent with our ARPES measurements, domains with multiple rotational alignments with respect to the underlying BLG are seen in the STM images of ML NbSe$_2$. Figs. S2a and S2b show STM images taken on domains with 14º and 22º angles between the NbSe$_2$ atomic lattice and the BLG lattice, respectively. The black arrows lie along the NbSe$_2$ lattice while the red arrows lie along the BLG lattice. Fig. S2c shows an atomically-resolved STM image of the bare BLG substrate, with the atomic lattice directions indicated by red arrows. The LEED pattern for a ML NbSe$_2$ film is shown in fig. S2d. The outer dots, circled in red, correspond to the BLG substrate and show well-defined rotational alignment across the sample. The features corresponding to the NbSe$_2$, however, appear as a smeared-out circle. This circle is not completely isotropic and displays increased intensity approximately every 30º. This is consistent with the ARPES measurements, which indicate two dominant rotational domains separated by 30º. The observed rotational disorder is a consequence of the weak interaction between the NbSe$_2$ and the underlying BLG substrate and has been observed in other monolayer TMD films grown on BLG.

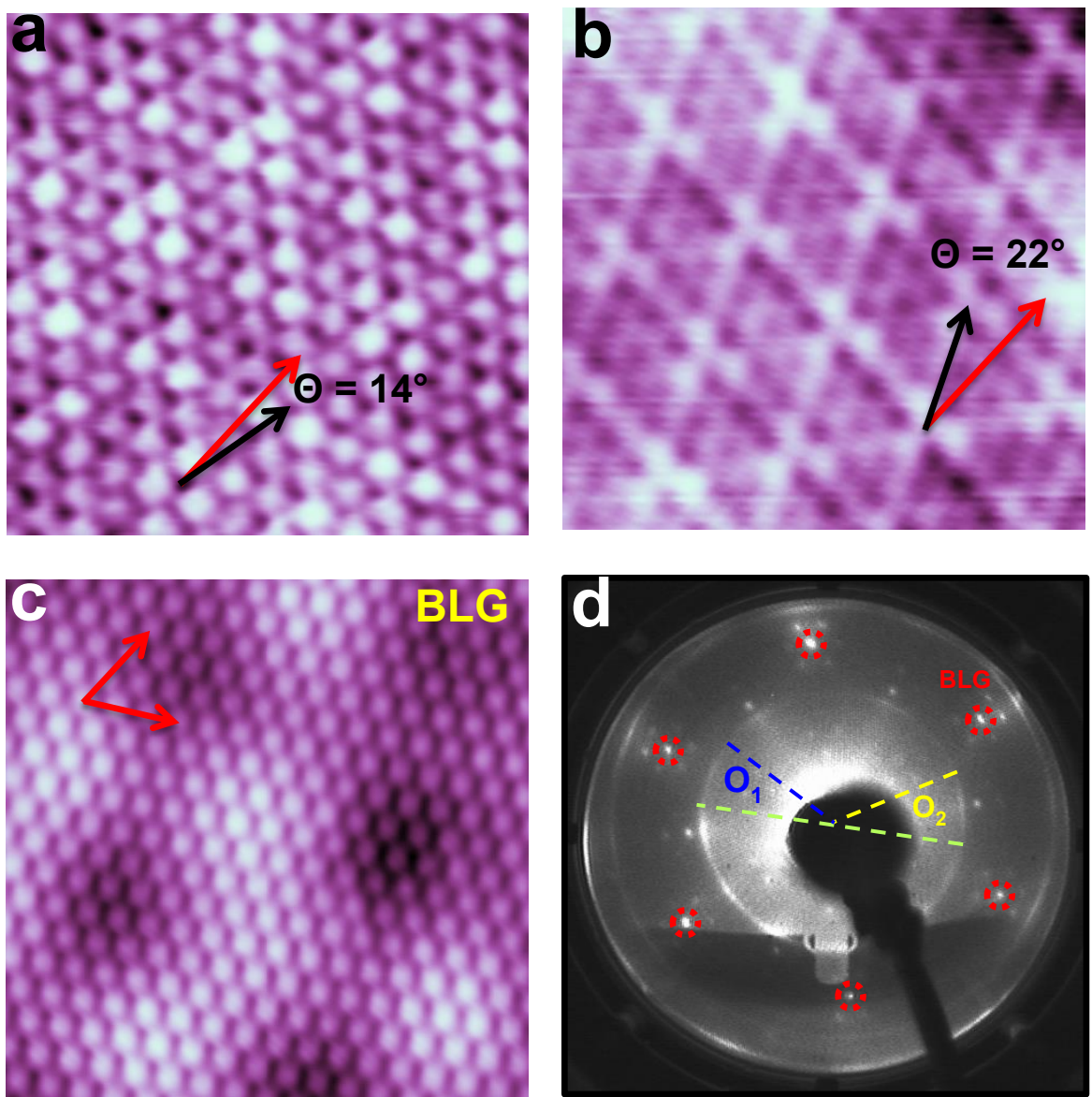

**Figure S2: Rotational orientation of single-layer NbSe₂ on BLG.** Atomically resolved STM images (40 x 40 Å²) of two domains of the NbSe₂ monolayer showing a rotational misalignment of the atomic lattice (black arrow) of **a.** 14° (Vs = - 20 mV, It = 1000 pA) and **b.** 22° (Vs = + 50 mV, It = 50 pA) with respect to the graphene lattice (red arrow). **c.** STM image of the BLG substrate indicating the orientation of the atomic lattice (40 x 40 Å², Vs = + 0.1 mV, It = 50 pA). **d.** LEED pattern of monolayer NbSe₂ showing the two preferred orientations of NbSe₂ ($O_1$ and $O_2$) and their large angular extension defined by the green-blue and green-yellow lines, respectively.

## 3. Transport measurements

The dc electrical resistance of the sample was measured using a 4-probe contact configuration as shown in Fig. S3.

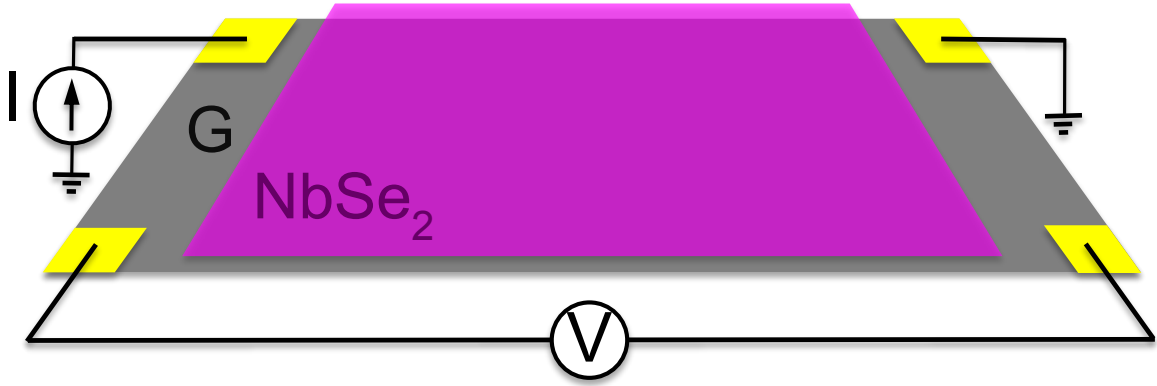

**Figure S3: Four-probe configuration for transport measurements.** Sketch of the 4-probe contact configuration used in the transport experiments. Graphene (G) layer is shown in grey and NbSe2 in purple. The insulating SiC(0001) substrate is not shown.

**References:**

1. Damascelli A., Hussain Z., Shen Z.-X. Angle-resolved photoemission studies of the cuprate superconductors. *Rev Mod Phys* 2003, **75,** 473-541.
2. Calandra M., Mazin I. I., Mauri F. Effect of dimensionality on the charge-density wave in few-layer *2H*-NbSe$_2$. *Phys Rev B* 2009, **80,** 241108.